\documentclass{article}
\usepackage{amssymb}
\usepackage{amsmath}
\usepackage{graphics}
\usepackage{epsf}
\usepackage{rotating}

\input{tcilatex}

\begin{document}

\title{Menelaus relation and Fay's trisecant formula are
associativity equations}
\author{B.G. Konopelchenko \\
%EndName
\\
Dipartimento di Fisica, Universita del Salento \\and INFN, Sezione
di Lecce, 73100 Lecce, Italy}

\maketitle

\begin{abstract}

 It is shown that the celebrated Menelaus relation and Fay's
trisecant formula similar to the WDVV equation are associativity
conditions for structure constants of certain three-dimensional
algebra.
\end{abstract}

\bigskip

\section{Introduction}

\ Associative algebras are fundamental ingredients in a number of
theories and constructions in theoretical and mathematical
physics. One of the most intriguing and unexpected \ recent
manifestation of their role is due to the discovery of Witten [1]
and Dijkgraaf-Verlinde-Verlinde [2]. They showed that the
properties of correlation functions $\langle \Phi _{j}\Phi
_{k}...\rangle $ for the two-dimensional topological field theory
are encoded by the algebraic relations of the form

\begin{equation}
\sum_{m=1}^{N}C_{jk}^{m}C_{ml}^{n}=%
\sum_{m=1}^{N}C_{kl}^{m}C_{jm}^{n},\quad j,k,l,n=1,...,N
\end{equation}
where $C_{kl}^{j}=\eta ^{jm}C_{mkl}$, $\eta ^{jm}$ is the matrix
inverse to the matrix of two-points correlation functions $\langle
\Phi _{j}\Phi _{k}\rangle $ and $C_{mkl}=\langle \Phi _{m}\Phi
_{k}\Phi _{l}\rangle $.
Moreover, for the deformed model the three-points correlation function $%
C_{mkl}(x)=\langle \Phi _{m}\Phi _{k}\Phi _{l}\rangle $ is the
third order derivative $C_{mkl}(x)=\frac{\partial ^{3}F}{\partial
x^{m}\partial x^{k}\partial x^{l}}$ and the algebraic equations
(1) take the form of the system of partial differential equations
for F (WDVV equation) [1,2].

A remarkable fact is that \ equations (1) are nothing else than
the associativity conditions for the structure constants of the
algebra of primary fields with the multiplication rule $\Phi
_{j}\cdot \Phi _{k}=\sum_{m=1}^{N}C_{jk}^{m}\Phi _{m}$ [1,2] and ,
hence, the WDVV\ equation is the associativity condition for
structure constants with a particular dependence on the
deformation parameters $x_{j}$.

\ This observation was beautifully formalized in [3,4] as the
theory of Frobenius manifolds and then extended to the theory of
F-manifolds in [5]. It turned out that the WDVV equation plays a
fundamental role in the theory of quantum cohomology and other
branches of algebraic geometry (see e.g. [4,6]). Thus, it was
demonstrated that the associativity equation (1) and its deformed
forms are fundamental objects encoding important information.

\ \ In this paper we will show that associativity equation plays
similar role in two other quite important cases, namely, for the
classical Menelaus relation and Fay's trisecant formula for
Riemann theta function. Separated in time by 2000 years and arosen
in a quite different branches of geometry both these formulas are
nothing but the associativity condition for a certain
three-dimensional algebra. The same is valid also for the bilinear
discrete Hirota-Miwa equation for the KP hierarchy.

The paper is organized as follows. In section 2 we briefly recall
the basic formulas for the simplest WDVV equation. Relation
between Menelaus configuration and theorem with associativity
condition is discussed in section 3. The KP case is considered in
section 4. The qauge equivalence of the Menelaus and KP
configurations is demonstrated in section 5. \ In section 6 it is
shown that Fay's trisecant formula also is the associativity
equation and a conjecture about the role of associative algebra in
characterization of Jacobian varieties is formulated.

\bigskip

\section{WDVV equation}

\ Here we will discuss briefly the simplest WDVV\ equation in order to
recall its connection with the associativity condition and in order to use
this construction further as a sort of guide. We will derive this equation
in a manner which is slightly different from the usual one ( see  [7,8]).

Thus, we consider a noncommutative three-dimensional associative algebra
\textit{A }with the unite element $\mathit{P}_{0}$ . We assume that the
algebra possess a commutative basis the elements of which we will denote as $%
\mathit{P}_{0},\mathit{P}_{1},\mathit{P}_{2}$. The table of multiplication $%
\mathit{P}_{0}\cdot \mathit{P}_{j}=\mathit{P}_{j},j=0,1,2$ and

\begin{eqnarray}
\mathbf{P}_{1}^{2} &=&A\mathbf{P}_{0}+B\mathbf{P}_{1}+C\mathbf{P}_{2}, \notag\\
\mathbf{P}_{1}\mathbf{P}_{2} &=&\mathbf{P}_{2}\mathbf{P}_{1}=D\mathbf{P}%
_{0}+E\mathbf{P}_{1}+G\mathbf{P}_{2}, \notag \\
\mathbf{P}_{2}^{2} &=&L\mathbf{P}_{0}+M\mathbf{P}_{1}+N\mathbf{P}_{2}
\end{eqnarray}
defines the structure constants A,B,C,...,N of the algebra
\textit{A} in
this basis. The associativity of the algebra, i.e. the conditions $(\mathbf{P%
}_{j}\mathbf{P}_{k})\mathbf{P}_{l}=\mathbf{P}_{j}(\mathbf{P}_{k}\mathbf{P}%
_{l}),j,k,l=0,1,2$ (conditions (1)) in this case are equivalent to the
following three equations

\begin{eqnarray}
A &=&EC+G^{2}-BG-CN,
\notag \\
D &=&CM-GE,
\notag \\
L &=&E^{2}+GM-MB-NE.
\end{eqnarray}

One of the ways to describe deformations of the structure constants
A,B,...,N is to associate the following system of linear differential
equations ( see e.g. [3], [8])

\begin{eqnarray}
\Psi _{x_{1}x_{1}} &=&A\Psi +B\Psi _{x_{1}}+C\Psi _{x_{2}},
\notag \\
\Psi _{x_{1}x_{2}} &=&D\Psi +E\Psi _{x_{1}}+G\Psi _{x_{2}},
\notag \\
\Psi _{x_{2}x_{2}} &=&L\Psi +M\Psi _{x_{1}}+N\Psi _{x_{2}}
\end{eqnarray}
with the multiplication table (2) \ (Dirac's recipe [8]) and
require its compatibility , i.e.

\begin{equation}
(\frac{\partial ^{2}}{\partial x_{j}\partial x_{k}})\frac{\partial \Psi }{%
\partial x_{l}}=\frac{\partial }{\partial x_{j}}(\frac{\partial ^{2}\Psi }{%
\partial x_{k}\partial x_{l}}).
\end{equation}
Here and below \ $\Psi _{x_{k}}=\frac{\partial \Psi }{\partial
x_{k}}$ etc. The corresponding system of nonlinear differential
equations for the structure constants admits various reductions.
One which we are interested in is $C=1,G=0,N=0$ . Under this
constraint the associativity conditions (3) are reduced to

\begin{equation}
A=E,\quad D=M
\end{equation}
and

\begin{equation}
L=A^{2}-DB
\end{equation}
while the system of differential equations becomes

\begin{eqnarray}
D_{x_{1}}-A_{x_{2}}+A^{2}-DB-L &=&0,
\notag \\
D_{x_{1}}-A_{x_{2}}+L+DB-A^{2} &=&0,
\notag \\
A_{x_{1}}-B_{x_{2}} &=&0,\quad L_{x_{1}}-D_{x_{2}}=0,
\notag \\
E-A &=&0,\quad M-D=0.
\end{eqnarray}
This system is equivalent to the algebraic associativity
conditions (6), (7) and differential exactness conditions

\begin{equation}
D_{x_{1}}-A_{x_{2}}=0,\quad A_{x_{1}}-B_{x_{2}}=0,\quad
L_{x_{1}}-D_{x_{2}}=0.
\end{equation}
Equations (9) imply the existence of a function F such that

\begin{eqnarray}
A =E=F_{x_{1}x_{1}x_{2}},\quad B=F_{x_{1}x_{1}x_{1}},
\notag \\
D =M=F_{x_{1}x_{2}x_{2}},\quad L=F_{x_{2}x_{2}x_{2}}.
\end{eqnarray}
The remaining associativity condition (7) thus becomes

\begin{equation}
F_{x_{2}x_{2}x_{2}}=(F_{x_{1}x_{1}x_{2}})^{2}-F_{x_{1}x_{1}x_{1}}F_{x_{1}x_{2}x_{2}}.
\end{equation}
It is the famous WDVV equation [1,2]. Its  algebro-geometrical
significance is discussed in [4,6].

We note that the WDVV\ equation (11) is nothing but the
associativity equation (7) in parametrization (10). \ Several
other integrable hydrodynamical type systems like the Hirota-Miwa
equation for dispersionless Kadomtsev-Petviashvili (KP) hierarchy
admit analogous algebraic interpretation [7].

\section{Menelaus relation as associativity condition}

\bigskip

\ In order to approach the Menelaus relation ( see e.g. [9,10]) in
a similar manner one should first choose an appropriate algebra.
Thus, we consider a noncommutative three-dimensional algebra
\textit{A}$_{M}$ without unite
element. We assume that it possess the commutative basis with the elements $%
\mathit{P}_{1},\mathit{P}_{2},\mathit{P}_{3}$ for which the products of only
distinct elements is defined. So, the multiplication table is

\begin{eqnarray}
\mathbf{P}_{1}\mathbf{P}_{2} =A\mathbf{P}_{1}+B\mathbf{P}_{2},
\notag \\
\mathbf{P}_{1}\mathbf{P}_{3} =C\mathbf{P}_{1}+D\mathbf{P}_{3},
\notag \\
\mathbf{P}_{2}\mathbf{P}_{3} =E\mathbf{P}_{2}+G\mathbf{P}_{3}.
\end{eqnarray}%
Associativity conditions

\begin{equation}
\mathbf{P}_{1}(\mathbf{P}_{2}\mathbf{P}_{3})=\mathbf{P}_{2}(\mathbf{P}_{3}%
\mathbf{P}_{1})=\mathbf{P}_{3}(\mathbf{P}_{1}\mathbf{P}_{2})
\end{equation}
for the structure constants of such algebra are

\begin{equation}
(A-G)C-EA=0,\quad (A-G)D+BG=0,\quad (C-E)B+DE=0
\end{equation}

\textbf{Lemma 1. }For nonzero A, B,...,G the associativity conditions (14)
are equivalent to the equation

\begin{equation}
AED+BCG=0
\end{equation}%
and one of equations (14), for instance, the equation

\begin{equation}
(A-G)C-EA=0.
\end{equation}

Proof. Multiplying the first of equations (14) by D, second by C
and subtracting results, one gets (15). The rest is
straightforward.

\ To describe deformations of the structure constants defined by (12) one
should, similar to the WDVV\ case, \ apply the Dirac's recipe to a linear
systems which will be realization of the table (12). \ We choose the
realization of $\mathit{P}_{1},\mathit{P}_{2},\mathit{P}_{3}$ by operators
of shifts $\mathit{P}_{j}=T_{j}$ where $T_{1}\Phi (n_{1},n_{2},n_{3})=\Phi
(n_{1}+1,n_{2},n_{3}),T_{2}\Phi (n_{1},n_{2},n_{3})=\Phi
(n_{1},n_{2}+1,n_{3}),T_{3}\Phi (n_{1},n_{2},n_{3})=\Phi
(n_{1},n_{2},n_{3}+1)$ and $n_{1},n_{2},n_{3}$ are deformation parameters [
11]. \ The corresponding linear system is [11 ]

\begin{equation}
\Phi _{12}=A\Phi _{1}+B\Phi _{2},\quad \Phi _{13}=C\Phi _{1}+D\Phi
_{3},\quad \Phi _{23}=E\Phi _{2}+G\Phi _{3}
\end{equation}
where $\Phi _{j}=T_{j}\Phi ,\Phi _{jk}=T_{j}T_{k}\Phi .$

Here we assume that all structure constants are real but $\Phi $ is a
complex-valued. \ So, $\Phi _{1},\Phi _{2},\Phi _{3},$ $\Phi _{12},\Phi
_{23},\Phi _{13}$ \ can be considered as points on the (complex) plane .
Thus, equations (17) with A,B,...,G obeying associativity conditions (15),
(16) define a configuration of six points on the plane.

There are at least two distinguished special configurations among them. The
first corresponds to the case when

\begin{equation}
A+B=1,\quad C+D=1,\quad E+G=1.
\end{equation}

For such A,B,...,G \ the relations (17), in virtue of the conditions (18),
mean that three points $\Phi _{1},\Phi _{2},\Phi _{12}$ are collinear as
well as the sets of points $\Phi _{1},\Phi _{3},\Phi _{13}$ and $\Phi
_{2},\Phi _{3},\Phi _{23}$. \ Then the relations (15), (16) imply that the
points $\Phi _{12},\Phi _{13},\Phi _{23}$ are collinear too, i.e.
\begin{equation}
\Phi _{12}=\frac{A}{C}\Phi _{13}+\frac{B}{E}\Phi _{23}
\end{equation}%
with $\frac{A}{C}+\frac{B}{E}=1$. \ Thus, in the case (18) the relations
(17) describe the set of four triples $(\Phi _{1},\Phi _{2},\Phi
_{12}),(\Phi _{1},\Phi _{3},\Phi _{13}),(\Phi _{2},\Phi _{3},\Phi _{23})$
and $(\Phi _{12},\Phi _{13},\Phi _{23})$ of collinear points. It is nothing
but the celebrated Menelaus configuration of the classical geometry (figure
1) ( see e.g. [90], [91]).
\begin{figure}[t]
\begin{center}
\includegraphics[width=8cm, angle=0]{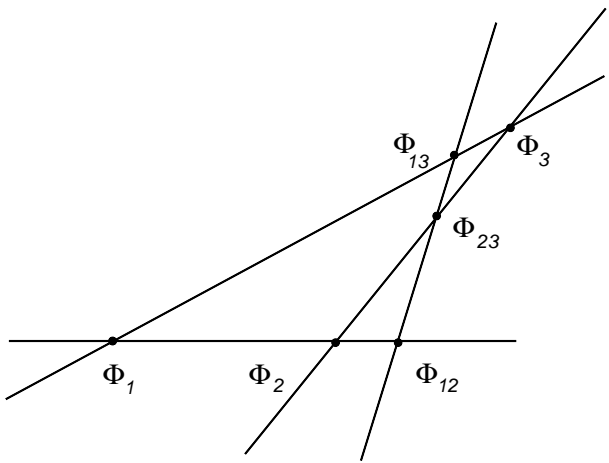}
\end{center}
\caption{{\protect\small Menelaus configuration.}}
\label{Figure1}
\end{figure}

The relations (17) and (18) allow us to express A,B,...,G in terms of $\Phi $%
. One gets

\begin{eqnarray}
A &=&\frac{\Phi _{12}^{M}-\Phi _{2}^{M}}{\Phi _{1}^{M}-\Phi _{2}^{M}},\quad
B=-\frac{\Phi _{12}^{M}-\Phi _{1}^{M}}{\Phi _{1}^{M}-\Phi _{2}^{M}},\quad C=%
\frac{\Phi _{13}^{M}-\Phi _{3}^{M}}{\Phi _{1}^{M}-\Phi _{3}^{M}},  \nonumber
\\
D &=&-\frac{\Phi _{13}^{M}-\Phi _{1}^{M}}{\Phi _{1}^{M}-\Phi _{3}^{M}},\quad
E=\frac{\Phi _{23}^{M}-\Phi _{3}^{M}}{\Phi _{2}^{M}-\Phi _{3}^{M}},\quad G=-%
\frac{\Phi _{23}^{M}-\Phi _{2}^{M}}{\Phi _{2}^{M}-\Phi _{3}^{M}}
\end{eqnarray}%
where we denote by $\Phi ^{M}$ solution of the system (17), (18). In such a
parametrization of A, B,...,G the associativity conditions (15), (16) are
equivalent to the single equation

\begin{equation}
\frac{(\Phi _{1}^{M}-\Phi _{12}^{M})(\Phi _{2}^{M}-\Phi _{23}^{M})(\Phi
_{3}^{M}-\Phi _{13}^{M})}{(\Phi _{12}^{M}-\Phi _{2}^{M})(\Phi _{23}^{M}-\Phi
_{3}^{M})(\Phi _{13}^{M}-\Phi _{1}^{M})}=-1.
\end{equation}
It is the celebrated  Menelaus relation (see [9], [10]) which is
necessary and sufficient condition for collinearity of the points
$\Phi _{12},\Phi _{13},\Phi _{23}$ \ for any three given points
$\Phi _{1},\Phi _{2},\Phi _{3} $ \ on the plane. In our
formulation the Menelaus relation (21) is nothing else than the
associativity conditions (15), (16) written in terms of $\Phi
^{M}$. Thus, the \ Menelaus theorem is intimately connected with
the associative algebra (12) with the choice (18).

\textbf{Proposition 1.} \ For  a configuration of six points on
the plane defined by equations (17) \ with the relation (18) the
following two conditions are equivalent:

(A) Coefficients A, B,...,G in (17) obey the associativity equations (14) or
(15), (16) for the algebra (12),

(B) Points $\Phi _{12},\Phi _{13},\Phi _{23}$ are collinear.

Proof. An implication (A)$\rightarrow $ (B) has been proved above. Now let
us assume that points $\Phi _{12},\Phi _{13},\Phi _{23}$ defined by (17)
together with (18) are collinear for arbitrary points $\Phi _{1},\Phi
_{2},\Phi _{3}$. This means that

\begin{equation}
\Phi _{12}=\alpha \Phi _{13}+\beta \Phi _{23}
\end{equation}
with $\alpha +\beta =1$. Substitution of expressions for $\Phi
_{12},\Phi _{13},\Phi _{23}$ given by (17), (18) into (22) \ gives
the \ conditions

\begin{equation*}
A=\alpha C,B=\beta E,\alpha D+\beta G=0.
\end{equation*}
These conditions imply that $AED+BCG=0$. Then the substitution of $\alpha =%
\frac{A}{C},\beta =\frac{B}{E}$ into the condition $A+B=1$ gives $AE+BC-CE=0$%
. It is easy to check that this equation is equivalent \ to the first
associtivity condition (14). Due to Lemma 1 these conditions are equivalent
to all associativity conditions (14) $\square $ .

Thus the Menelaus configuration and theorem are just the geometric
realizations of the associativity conditions for the algebra (12).

\ Discrete deformations of the Menelaus configurations are
governed by discrete equations arising as compatibility conditions

\begin{equation*}
\Phi _{3(12)}=\Phi _{1(23)}=\Phi _{2(13)}
\end{equation*}
for the system (17). They are of the form

\begin{equation}
\frac{A_{3}}{A}=\frac{C_{2}}{C},\qquad \frac{B_{3}}{B}=\frac{E_{1}}{E}%
,\qquad \frac{D_{2}}{D}=\frac{G_{1}}{G},
\end{equation}%
\begin{equation}
(A_{3}-G_{1})C-E_{1}A=0,
\end{equation}%
\begin{equation}
(A_{3}-G_{1})D+B_{3}G=0,
\end{equation}%
\begin{equation}
(C_{2}-E_{1})B+D_{2}E=0.
\end{equation}%
Equations (24), (25) imply that

\begin{equation}
AE_{1}D+B_{3}CG=0
\end{equation}%
which is equivalent to the relation (15) in virtue of the second
equation (23). Thus, the Menelaus relation (21) is preserved by
deformations. Discrete deformations of the Menelaus configuration
given by equations (23)-(26) generate an integrable lattice on the
plane \ [12].

\bigskip

\section{KP configurations, discrete KP deformations and
Hirota-Miwa equation}

\bigskip

Another distinguished case corresponds to the choice

\begin{equation}
A+B=0,\quad C+D=0,\quad E+G=0
\end{equation}
for which equations (17) take the form

\begin{equation}
\Phi _{12}=A(\Phi _{1}-\Phi _{2}),\quad \Phi _{13}=C(\Phi
_{1}-\Phi _{3}),\quad \Phi _{23}=E(\Phi _{2}-\Phi _{3}).
\end{equation}
With such a choice of A,B,...,G the relation (15) is a trivial
identity and, hence, the associativity conditions are reduced to
the single equation

\begin{equation}
AC+EC-AE=0.
\end{equation}

\ Geometrical configuration on the plane formed by six points $\Phi
_{1},\Phi _{2},\Phi _{3},$ $\Phi _{12},\Phi _{23},\Phi _{13}$ with real A,
C, E is a special one. We, first, observe that the points $\Phi _{12},\Phi
_{13},\Phi _{23}$ lie on the straight lines passing through the origin 0 and
parallel to the straight lines passing through the points ($\Phi _{1},\Phi
_{2}),(\Phi _{1},\Phi _{3}),$ $(\Phi _{2},\Phi _{3})$, respectively. Then,
due to the associativity condition (30) the points $\Phi _{12},\Phi
_{23},\Phi _{13}$ are collinear. Indeed, equations (29) imply that

\begin{equation}
\frac{1}{C}\Phi _{13}-\frac{1}{A}\Phi _{12}-\frac{1}{E}\Phi _{23}=0
\end{equation}
while the relation (30) is equivalent to the condition $\frac{1}{C}-\frac{1}{%
A}-\frac{1}{E}=0.$Thus, the points $\Phi _{1},\Phi _{2},\Phi _{3},$ $\Phi
_{12},\Phi _{23},\Phi _{13}$ form the configuration on the complex plane
shown on the figure 2.

\begin{figure}[tbp]
\begin{center}
\includegraphics[width=12cm, angle=0]{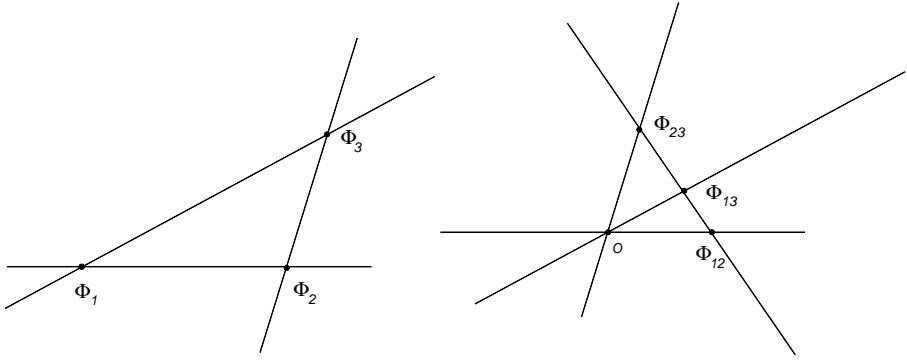}
\end{center}
\caption{{\protect\small $KP$ configuration.}}
\label{Figure2}
\end{figure}

\ The associativity condition (30) provides us also with the relation
between the directed lengths for this configuration. Indeed, expressing A,
C, E from (29) in terms of $\Phi $ and substituting into (30), one gets

\begin{equation}
\frac{\Phi _{1}-\Phi _{2}}{\Phi _{12}}+\frac{\Phi _{2}-\Phi _{3}}{\Phi _{23}}%
+\frac{\Phi _{3}-\Phi _{1}}{\Phi _{31}}=0.
\end{equation}%
Since for real A, C, E $\frac{\Phi _{1}-\Phi _{2}}{\Phi _{12}}=\frac{\left|
\Phi _{1}-\Phi _{2}\right| }{\left| \Phi _{12}\right| }$ etc the formula
(32) represents the relation between the directed lengths $\left| \Phi
_{1}-\Phi _{2}\right| $ of the interval $(\Phi _{1},\Phi _{2})$ etc. \

We note that the straight line passing through the points $\Phi _{12},\Phi
_{23},\Phi _{13}$ is a trisecant of the family of three straight lines
passing through origin. We will refer to the configuration presented in the
figure 2 as KP configuration by the reason which will be clarified now.

\ \ Let us consider discrete deformations of such configurations. They are
governed by equations (23)-(26) under the constraint (28).

\textbf{Lemma 2.} In the case (28) equations (23)-(26) are
equivalent to the associativity condition (30) and equations

\begin{equation}
\frac{A_{3}}{A}=\frac{C_{2}}{C}=\frac{E_{1}}{E}.
\end{equation}

Proof. In this case equations (23) are reduced to equations (33) while
equations (24)-(26) are equivalent to the single equation $%
A_{3}C+E_{1}C-AE_{1}=0$ . Due to (33) this \ equation is equivalent to the
associativity condition (30)$\square $.

 Equations (33) imply the existence of the function $\tau $ such that

\begin{equation}
A=-\frac{\tau _{1}\tau _{2}}{\tau \tau _{12}},\quad C=-\frac{\tau
_{1}\tau _{3}}{\tau \tau _{13}},\quad E=-\frac{\tau _{2}\tau
_{3}}{\tau \tau _{23}}.
\end{equation}
Substitution of these expressions into (30) gives

\begin{equation}
\tau _{1}\tau _{23}-\tau _{2}\tau _{13}+\tau _{3}\tau _{12}=0
\end{equation}
which is the celebrated discrete bilinear Hirota-Miwa equation for
the KP hierarchy or the addition formula for KP $\tau -$function
[13]. This fact justifies the name of the configuration. We would
like to emphasize that the Hirota-Miwa equation (35) is nothing
but the associativity condition (30) with the structure constants
A, C, E parametrized by $\tau $-function. We note that equations
(29) with A, C, E of the form (34) coincide with well-known linear
problems for the Hirota-Miwa bilinear equation [13].

Similar to the Menelaus case we thus have

\textbf{Proposition 2. }\ For the six-points configuration on the plane
defined by equations (28), (29) the following conditions are equivalent:

(A) Coefficients A,E,C obey the associativity condition (30),

(B) The points $\Phi _{12},\Phi _{23},\Phi _{13}$ are collinear,

(C) Function $\tau $ defined by (34) obeys the discrete Hirota-Miwa equation
(35).

\ Proof. Implications (A)$\rightarrow $(B) and (A)$\rightarrow $ (C) have
been proved above. Implication (B)$\rightarrow $ (A) is obvious. Then,
multiplying equation (35) by $\frac{\tau }{\tau _{1}\tau _{2}\tau _{3}}$,
one gets (30) that proves (C)$\rightarrow $ (A) and, hence, (C)$\rightarrow $
(B) $\square .$

\bigskip

\section{Gauge equivalence of the Menelaus and KP configurations}

\bigskip

The Menelaus and KP configurations look quite different. For instance, the
points $\Phi _{1},\Phi _{2},\Phi _{12}$ etc are collinear in the Menelaus
case and they are not in the KP case. Nevertheless, they are closely
connected, namely, they are gauge equivalent to each other.

\ \ A notion of gauge equivalency is a very well-known one in the theory of
gauge fields as well in the \ theory of integrable equations. \ In our case
the linear system (17) is invariant under the gauge transformations $\Phi
\rightarrow \widetilde{\Phi }=g^{-1}\Phi $ and

\begin{eqnarray}
\widetilde{A} &=&\frac{g_{1}}{g_{12}}A^{{}},\quad \widetilde{B}=\frac{g_{2}}{%
g_{12}}B,\quad \widetilde{C}=\frac{g_{1}}{g_{13}}C,  \nonumber \\
\widetilde{D} &=&\frac{g_{3}}{g_{13}}D,\quad \widetilde{E}=\frac{g_{2}}{%
g_{23}}E,\quad \widetilde{G}=\frac{g_{3}}{g_{23}}G
\end{eqnarray}
where $g(n_{1},n_{2},n_{3})$ is an arbitrary function. It is a
simple check \ that the system (23)-(26) is invariant under these
gauge transformations too.

\textbf{Lemma 3} [11]\textbf{\ }\ The relation (15) is invariant under the
gauge transformations (36).

Proof. Indeed, under the gauge transformation (36) one has

\begin{equation}
\widetilde{A}\widetilde{E}\widetilde{D}+\widetilde{B}\widetilde{C}\widetilde{%
G}=\frac{g_{1}g_{2}g_{3}}{g_{12}g_{23}g_{13}}(AED+BCG)\square
\end{equation}
So the relation (15) is a characteristic one for orbits of gauge
equivalent structure constants.

Gauge invariance of the general system (23)-(26) allows us to choose
different gauges. First , we observe that

\begin{equation}
\widetilde{A}+\widetilde{B}=\frac{g_{1}A+g_{2}B}{g_{12}},\quad \widetilde{C}+%
\widetilde{D}=\frac{g_{1}C+g_{3}D}{g_{13}},\quad \widetilde{E}+\widetilde{G}=%
\frac{g_{2}E+g_{3}G}{g_{23}}.
\end{equation}
So, if for generic A,B,...,G one chooses the gauge function g to
be a solution $\widehat{\Phi }$ of the linear system (17) then the
gauge transformed $\widetilde{A},\widetilde{B},...,\widetilde{G}$
obey the relations

\begin{equation}
\widetilde{A}+\widetilde{B}=1,\quad
\widetilde{C}+\widetilde{D}=1,\quad \widetilde{E}+\widetilde{G}=1.
\end{equation}

Thus, the relation (18) discussed in section (3) selects a special gauge
which is nothing but the distinguished Menelaus gauge.

Solutions of the linear system (17) in this gauge are ratios of two
solutions of the generic system (17) : $\widehat{\Phi }=\frac{\Phi }{%
\widetilde{\Phi }}$. So, if one treats \ solutions $\Phi $ of the
general system (17) as projective homogeneous coordinates , then
in terms of affine coordinates $\frac{\Phi }{\widetilde{\Phi }}$
it is the Menelaus system (system (17) with the relations (18) ).
In other words, the Menelaus configuration is the affine form of
the generic configuration of six points defined by the system
(17). We note that the gauge transformation (36) geometrically
means a local (depending on a point) homothetic transformation.

\ Now let us begin with the particular KP system (17), i.e. when
A,B,...,G obey the relations (28). \ In this case after the gauge
transformation one has the relations (38) with B=-A, D=-C, G=-E. \
Choosing g as a solution of the KP system (29), one again gets
(39).

\ So, KP and Menelaus cases are qauge equivalent to each other and $\Phi
^{M}=\frac{\widetilde{\Phi }^{KP}}{\widehat{\Phi }^{KP}}$ where $\widetilde{%
\Phi }^{KP},\widehat{\Phi }^{KP}$ are two distinct solutions of the system
(29). Hence, in order to construct the Menelaus configuration one needs two
KP configurations with the same A,C,E.

 If one starts with the Menelaus gauge then in order to get the relations $%
\widetilde{A}+\widetilde{B}=0,\quad
\widetilde{C}+\widetilde{D}=0,\quad \widetilde{E}+\widetilde{G}=0$
one should consider the gauge transformation with g obeying the
equations

\begin{equation*}
g_{1}A+g_{2}B=0,\quad g_{1}C+g_{3}D=0,\quad g_{2}E+g_{3}G=0.
\end{equation*}
It is a straightforward check that these equations are compatible
due to the relations (15) and equations (23)-(26). The same
arguments shows that there exists a gauge transformation which
converts the general system (17) into the KP linear system.

\ Thus, the Menelaus and KP cases and configurations are two particular, but
distinguished members of the orbit of structure constants generated by gauge
transformations (36).

\ Menelaus and KP gauges are distinguished also from the point of
view of collinearity property for \ the points $\Phi _{12},\Phi
_{23},\Phi _{13}$ . \ Indeed, let us consider a family of
configuration of six points defined by the system (17) for which
points $\Phi _{12},\Phi _{23},\Phi _{13}$ are collinear, i.e.
$\alpha \Phi _{12}+\beta \Phi _{23}+\gamma \Phi _{13}=0$ with
$\alpha +\beta +\gamma =1$. It is easy to show that these
conditions are satisfied if A,B,...,G obey the constraints
$AED+BCG=0$ and

\begin{equation}
A+B=C+D=E+G.
\end{equation}

Due to the gauge freedom and admissibility of rescaling
$A\rightarrow \lambda A,B\rightarrow \lambda B,...,G\rightarrow
\lambda G$ the condition (40) is gauge equivalent to the following
$A+B=C+D=E+G=\mu $ where $\mu $ is  an arbitrary constant. The
cases $\mu \neq 0$ are all
equivalent to the Menelaus case $\mu =1$ by rescaling of A,...,G. The case $%
\mu =0$ corresponds to the KP gauge.

\bigskip

\section{Fay's trisecant formula and associativity}

\bigskip

 The existence of the solutions for the KP $\tau -$function in terms of the
Riemann theta function (see e.g.[14]) and well-known relation
between the addition formula for KP \ hierarchy and Fay's
trisecant formula (see e.g. [15,16]) suggest that the latter also
is deeply connected with associativity. We will show here that it
is indeed the case. \

\ We recall the Fay's trisecant formula in its original form (see [15]). Let
X is a Riemann surface of genus g, $\theta (z)$ is an associated
theta-function ($z\in C^{g})$, E(x,y) is a prime form, $\omega =\{\omega
_{i},i=1,..,g)$ is a basis of holomorphic one-forms on X, $\widetilde{X}$ is
a universal covering of X and arbitrary $\alpha _{0},\alpha _{1},\alpha
_{2},\alpha _{3}\in \widetilde{X}$ . The Fay's trisecant formula is [15]

\begin{eqnarray}
\theta (z+\int_{\alpha _{0}}^{\alpha _{2}}\omega )\theta
(z+\int_{\alpha _{1}}^{\alpha _{3}}\omega )E(\alpha _{0},\alpha
_{3})E(\alpha _{2},\alpha _{1})+
\notag \\
\theta (z+\int_{\alpha
_{1}}^{\alpha _{2}}\omega )\theta (z+\int_{\alpha _{0}}^{\alpha
_{3}}\omega )E(\alpha _{2},\alpha _{0})E(\alpha _{3},\alpha _{1})-
\notag \\
 \theta (z)\theta (z+\int_{\alpha _{0}+\alpha _{1}}^{\alpha
_{2}+\alpha _{3}}\omega )E(\alpha _{0},\alpha _{1})E(\alpha
_{2},\alpha _{3})=0.
\end{eqnarray}
Let us rewrite this formula in the equivalent form. Using the
identity (see e.g. [15,16])

\begin{equation}
\int_{\alpha _{0}}^{\alpha _{2}}\omega +\int_{\alpha _{1}}^{\alpha
_{3}}\omega =\int_{\alpha _{1}}^{\alpha _{2}}\omega +\int_{\alpha
_{0}}^{\alpha _{3}}\omega =\int_{\alpha _{0}+\alpha _{1}}^{\alpha
_{2}+\alpha _{3}}\omega ,
\end{equation}
one gets

\begin{eqnarray}
\theta (z+\int_{\alpha _{0}}^{\alpha _{2}}\omega )\theta
(z+\int_{\alpha _{0}}^{\alpha _{3}}\omega -\int_{\alpha
_{0}}^{\alpha _{1}}\omega )E(\alpha
_{0},\alpha _{3})E(\alpha _{2},\alpha _{1})+
\notag \\
+\theta (z+\int_{\alpha _{0}}^{\alpha _{2}}\omega -\int_{\alpha
_{0}}^{\alpha _{1}}\omega )\theta (z+\int_{\alpha _{0}}^{\alpha
_{3}}\omega
)E(\alpha _{2},\alpha _{0})E(\alpha _{3},\alpha _{1})-
\notag \\
-\theta (z)\theta (z+\int_{\alpha _{0}}^{\alpha _{2}}\omega +\int_{\alpha
_{0}}^{\alpha _{3}}\omega -\int_{\alpha _{0}}^{\alpha _{1}}\omega )E(\alpha
_{0},\alpha _{1})E(\alpha _{2},\alpha _{3}) &=&0.
\end{eqnarray}
Then we fix all $\alpha _{j}$, denote $U_{j}=\int_{\alpha
_{0}}^{\alpha _{j}}\omega ,$ $(\alpha _{1}\neq \alpha _{2}\neq
\alpha _{3}\neq \alpha _{1}) $ and introduce shifts $T_{j}$
defined by

\begin{equation}
T_{j}\theta (z)=\theta (z+U_{j}),\quad j=1,2,3.
\end{equation}
Note that $U_{j}$ are commonly used objects connected with the
Abel's map ( see e.g. [14]). \ In these notations the Fay's
formula \ (43) takes the form

\begin{equation}
aT_{2}\theta (z)\cdot T_{1}^{-1}T_{3}\theta
(z)+bT_{1}^{-1}T_{2}\theta (z)\cdot T_{3}\theta (z)+c\theta
(z)\cdot T_{1}^{-1}T_{2}T_{3}\theta (z)=0
\end{equation}
where $a=E(\alpha _{0},\alpha _{3})E(\alpha _{2},\alpha
_{1}),b=E(\alpha _{2},\alpha _{0})E(\alpha _{3},\alpha
_{1}),c=-E(\alpha _{0},\alpha _{1})E(\alpha _{2},\alpha _{3}).$
Applying the shift $T_{1}$ to the l.h.s. of (45), one finally gets

\begin{equation}
aT_{3}\theta (z)\cdot T_{1}T_{2}\theta (z)+bT_{2}\theta (z)\cdot
T_{1}T_{3}\theta (z)+cT_{1}\theta (z)\cdot T_{2}T_{3}\theta (z)=0.
\end{equation}
Comparing (46) with (35), one readily recognizes in (46) the
associativity condition (30) with

\begin{equation}
A=\frac{1}{a}\frac{T_{1}\theta \cdot T_{2}\theta }{\theta \cdot
T_{1}T_{2}\theta },\quad C=-\frac{1}{b}\frac{T_{1}\theta \cdot T_{3}\theta }{%
\theta \cdot T_{1}T_{3}\theta },\quad
E=\frac{1}{c}\frac{T_{2}\theta \cdot T_{3}\theta }{\theta \cdot
T_{2}T_{3}\theta }.
\end{equation}
Then, as in the KP case one considers the system of linear
equations

\begin{eqnarray}
T_{1}T_{2}\Phi (z) &=&A(T_{1}\Phi (z)-T_{2}\Phi (z)),
\notag \\
T_{1}T_{3}\Phi (z) &=&C(T_{1}\Phi (z)-T_{3}\Phi (z)),
\notag \\
T_{2}T_{3}\Phi (z) &=&E(T_{2}\Phi (z)-T_{3}\Phi (z)).
\end{eqnarray}
This system implies that

\begin{equation}
\frac{1}{A}T_{1}T_{2}\Phi (z)+\frac{1}{E}T_{2}T_{3}\Phi (z)-\frac{1}{C}%
T_{1}T_{3}\Phi (z)=0.
\end{equation}
The associativity condition (30) means that the points $\Phi
(z+U_{1}+U_{2}),\Phi (z+U_{2}+U_{3}),\Phi (z+U_{1}+U_{3})$ are
collinear. Thus, we have

\textbf{\ Proposition 3.} The following three conditions are equivalent:

(A) Function $\theta (z)$ obeys the Fay's trisecant formula (41).

(B) Structure constants A,C,E defined by (47) obey the associativity
condition

\begin{equation}
AC+EC-AE=0,
\end{equation}
for the algebra (12) in the gauge $A+B=C+D=E+G=0,$

(C) Three points $\Phi (z+U_{1}+U_{2}),\Phi (z+U_{1}+U_{3}),\Phi
(z+U_{2}+U_{3})$ defined by relations (48) are collinear.

Proof. Equivalence of (A) and (B) has been proved above.
Equivalence of (B) and (C) is obvious.$\square$

\ Thus, the Fay's trisecant formula for theta-function of any
Riemann surface is nothing but the associativity condition for the
structure constants of the algebra (12) with parametrization (47).
The latter is a consequence of deformation equations (33). We
emphasize also that the collinearity of the points $\Phi
(z+U_{1}+U_{2}),\Phi (z+U_{1}+U_{3}),\Phi (z+U_{2}+U_{3})$ is
equivalent to the associativity condition.

Applying
$T_{1}^{-\frac{1}{2}}T_{2}^{-\frac{1}{2}}T_{3}^{-\frac{1}{2}}$ to
the relation (49), one gets

\begin{equation}
\frac{1}{\widetilde{A}}T_{1}^{\frac{1}{2}}T_{2}^{\frac{1}{2}}T_{3}^{-\frac{1%
}{2}}\Phi +\frac{1}{\widetilde{E}}T_{1}^{-\frac{1}{2}}T_{2}^{\frac{1}{2}%
}T_{3}^{\frac{1}{2}}\Phi -\frac{1}{\widetilde{C}}T_{1}^{\frac{1}{2}}T_{2}^{-%
\frac{1}{2}}T_{3}^{\frac{1}{2}}\Phi =0
\end{equation}
where

\begin{equation}
\frac{1}{\widetilde{A}}+\frac{1}{\widetilde{E}}-\frac{1}{\widetilde{C}}=0.
\end{equation}

Thus, the points

\begin{equation}
\Phi (z+\frac{1}{2}(U_{1}+U_{2}-U_{3})),\quad \Phi (z+\frac{1}{2}%
(-U_{1}+U_{2}+U_{3})),\quad \Phi
(z+\frac{1}{2}(U_{1}-U_{2}+U_{3}))
\end{equation}
are collinear too. Standard Fay's trisecant formula (41) is
equivalent ( see e.g. [15,16]) to the collinearity of the points

\begin{equation}
\varphi (z+\frac{1}{2}(U_{1}+U_{2}-U_{3})),\quad \varphi (z+\frac{1}{2}%
(-U_{1}+U_{2}+U_{3})),\quad \varphi
(z+\frac{1}{2}(U_{1}-U_{2}+U_{3}))
\end{equation}
in the Kummer variety. It suggests to identify the map $\Phi $
defined by (48) with the Kummer map: $\Phi =\varphi =\theta \left[
\begin{array}{c}
0 \\
\eta%
\end{array}%
\right] (z,\frac{\Omega }{2})_{\eta \in \frac{1}{2}Z^{g}/Z^{g}}.$

In the papers [17-21] is was shown that the existence of a family
of trisecants or even only one trisecant characterize Jacobian
varieties among indecomposable principally polarized abelian
varieties (ppavs). The results of the Proposition 3 make it quite
natural to \textbf{conjecture }that the existence of the
associative three-dimensional algebra (12), (28) for ppav such
that on the corresponding line bundle equations (48) are valid
with shifts $T_{j}$ defined by (44) is characteristic one for
Jacobian varieties.

Equations (46), (48), (49) can be treated as discrete equations of one takes
theta function and function $\Phi $ of the form $\theta
(n_{1},n_{2},n_{3};z)=\theta (z+\sum_{1}^{3}U_{i}n_{i})$ and $\Phi
(n_{1},n_{2},n_{3};z)=\Phi (z+\sum_{1}^{3}U_{i}n_{i})$ \ (see e.g. [14] for
the theta-function solutions of the Hirota-Miwa equation (35) ). These
equations admit different reductions. For instance, if one frozes the
dependence on $n_{3}$ then under the constraint $\Phi =\frac{\theta
(A+U_{1}n_{1}+U_{2}n_{2}+z)}{\theta (U_{1}n_{1}+U_{2}n_{2}+z)}$ $\exp
(n_{1}\alpha +n_{2}\beta )$ where fixed $A=U_{3}n_{3}$ and $\alpha ,\beta $
are constants the first equation (48) coincides with equation (1.14) (
together with (15), (16) ) from the paper [21] which is necessary and
sufficient condition for ppav to be Jacobian of a smooth curve of genus g. \
This fact supports the \textbf{Conjecture }formulated above.

\bigskip

\textbf{Acknowlegment.} The author thanks A.Veselov and A.
Nakayashiki for useful and stimulating discussions.

\bigskip

\textbf{References}

[1] Witten E.,  On the structure of topological phase of
two-dimensional gravity \textit{Nucl.Phys.} \textbf{B 340} 281-332
(1990)

[2] Dijkgraaf R, Verlinde H and Verlinde E  Topological strings in
d<1 , \textit{Nucl.Phys.} \textbf{\ B 352 }59-86 (1991)

[3] Dubrovin B  Integrable systems in topological field theory \textit{%
Nucl.Phys.} \textbf{B 379} 627-689 (1992)

[4] Dubrovin B.  \ Geometry of 2D topological field theories, Lecture Notes
in Math. \textbf{1620}, 120-348 (Berlin: Springer-Verlag), 1996

[5] Hertling C and Manin Y I  Weak Frobenius manifolds \textit{%
Int.Math.Res.Notices } \textbf{6} 277-286 (1999)

[6] Manin Y I 1999 \textit{Frobenius manifolds, quantum cohomology and
moduli spaces} (Providence: AMS) 1999.

[7] Konopelchenko B G and \ Magri F \ Coisotropic deformations of
associative algebras and dispersionless integrable hierarchies, \textit{%
Commun. Math. Phys. } \textbf{274} 627-658 (2007)

[8] Konopelchenko B G  Quantum deformations of associative algebras and
integrable systems, \textit{J.Phys.A: Math. Theor. }\textbf{42 }095201 (2009)

[9]  Pedoe D  \textit{Geometry. A comprehensive course} ( Cambridge:
Cambridge University Press), 1970

[10] \ Brannan D A, Esplen M F and Gray J \textit{Geometry} ( Cambridge:
Cambridge University Press), 1999

[11] \ Konopelchenko B.G., Discrete integrable systems and deformations of
associative algebras, \ arXiv:0904.2284, 2009

[12]  Konopelchenko B G and \ Schief W K , Menelaus' theorem, Clifford
configurations and inversive geometry of the Schwarzian KP hierarchy,
\textit{J. Phys. A: Math. Gen. }\textbf{35 }6125-6144 (2002)

[13] Miwa T., On Hirota's difference equation, \textit{Proc.Japan Acad.}%
\textbf{\ 58A }8-11 (1982)

[14] Krichever I., Wiegmann P. and Zabrodin A., Elliptic solutions to
difference non-linear equations and related many-body problems, Commun.
Math. Phys., \textbf{193}, 373-396 (1998).

[15] \ Fay J. D., Theta functions on Riemann surfaces, Lecture Notes in
Mathematics, vol \textbf{352}, Berlin, Springer-Verlag, 1973

[16] \ Mumford D., Tata lectures on Theta I, II, Boston, Birkhauser,
1983,1984

[17] \ Gunning R., \ Some curves in abelian varieties, \ Invent. Math.,
\textbf{66}, 377-389 (1982)

[18]  Welters G. E., \ A criterion for Jacobian varieties, Ann. Math.,
\textbf{120}, 497-504 (1984)

[19] \ Debarre O., \ Trisecant lines and Jacobians, II, Compisito
Math. \textbf{107}, 177-186 (1997).

[20] Arbarello E., Krichever I. and Marini G., Characterizing Jacobians via
flexes of the Kummer variety, Math. Res. Lett., \textbf{13:}1 , 109-123
(2006).

[21] \ Krichever I., Characterizing Jacobians via trisecant of the
Kummer variety, Ann. of Math.; arXiv:math/0605625 (2008).

\end{document}